\documentclass[aps,twocolumn]{revtex4}

\RequirePackage[T2A]{fontenc}

\usepackage{amssymb}
\usepackage{amsmath}
\usepackage{bm}
\usepackage[dvips]{epsfig}
\usepackage{graphicx}
\usepackage{epigraph}

\begin{document}

\title{Open level lines of a superposition of periodic 
potentials on a plane}

\epigraph{Dedicated to the 90th anniversary of  \linebreak
I.E. Dzyaloshinskii}

\author{A.Ya. Maltsev$^{1}$, S.P. Novikov$^{1,2}$}

\affiliation{
\centerline{$^{1}$ \it{L.D. Landau Institute for Theoretical Physics 
of Russian Academy of Sciences}}
\centerline{\it 142432 Chernogolovka, pr. Ak. Semenova 1A}
\centerline{$^{2}$ \it{V.A. Steklov Mathematical Institute 
of Russian Academy of Sciences}}
\centerline{\it 119991 Moscow, Gubkina str. 8}
}

\begin{abstract}
 We consider here open level lines of potentials resulting from 
the superposition of two different periodic potentials on the plane.
This problem can be considered as a particular case of the Novikov 
problem on the behavior of open level lines of quasi-periodic 
potentials on the plane with four quasi-periods. At the same time, 
the formulation of this problem may have many additional features 
that arise in important physical systems related to it. Here we 
will try to give a general description of the emerging picture 
both in the most general case and in the presence of additional 
restrictions. The main approach to describing the possible behavior 
of the open level lines will be based on their division into 
topologically regular and chaotic level lines.
\end{abstract}

\maketitle

\section{Introduction}

 In this paper, we consider applications of the theory 
of quasi-periodic functions to two-dimensional systems 
formed by a superposition of periodic systems on a plane. 
More precisely, we will consider applications of the latest 
results obtained in the study of the Novikov problem 
with four quasi-periods to potentials generated 
by a superposition of two arbitrary periodic potentials
with their possible influence on each other.

 The beginning of the theory of quasi-periodic functions 
goes back to the works of H. Bohr and A.S. Besikovich 
(see e.g. \cite{Bohr,Besicovitch}). Generally speaking, 
there are a number of definitions of a quasi-periodic 
function in the literature. Here we will call a quasi-periodic 
function $f (x^{1}, \dots , x^{n}) $ in $\mathbb{R}^{n}$ 
the restriction of a periodic function 
$F (z^{1}, \dots , z^{N}) $, defined in space 
$\mathbb{R}^{N}$, under an affine embedding
$\mathbb{R}^{n} \rightarrow \mathbb{R}^{N}$.
The dimension $N$ of the embedding space is then called 
the number of quasi-periods of the function
$f (x^{1}, \dots , x^{n}) $. Let us also note here,
that this definition is well known, for instance, in the
theory of quasicrystals.

 The general problem of Novikov is to describe the 
geometry of the level lines of quasi-periodic functions on the 
plane with an arbitrary number of quasi-periods. It must be said 
that this problem was originally set for functions with three 
quasi-periods (\cite{MultValAnMorseTheory}). The problem 
in this formulation is directly related to the theory of 
galvanomagnetic phenomena in metals with complex Fermi surfaces.
The Novikov problem with three quasi-periods has now been studied 
in most detail. In particular, a complete classification of all 
types of level lines of the corresponding functions on the plane 
has been obtained, and a number of important consequences of such 
a classification for the theory of transport phenomena have been 
described. Here, however, we will not consider the case of three 
quasi-periods and give just some references to reviews on this topic
(see e.g. \cite{UmnObzor,ObzorJetp}).

 In this paper, the central role will be played by the Novikov 
problem with four quasi-periods. The case of four quasi-periods 
has not been studied in as much detail as the case of three 
quasi-periods, however, for this case there are also deep analytical 
results (\cite{NovKvazFunc,DynNov}). As can be shown, the potentials 
formed by the superposition of two periodic potentials in the plane 
are potentials with four quasi-periods, and the results obtained in 
the works \cite{NovKvazFunc,DynNov} can thus be applied to them.
It should be noted, however, that the results of the papers 
\cite{NovKvazFunc,DynNov} are formulated in terms of the original 
parameter space of the problem (i.e., parameters describing the 
embedding of a two-dimensional plane in the space $\mathbb{R}^{4}$)
that differs from the space of parameters we are considering here. 
Therefore, we will have to specifically consider here the connection
between the initial parameter space and the parameters of the problem 
we are studying. As we will see, this connection strongly depends 
in fact on the specifics of the problem under consideration, which 
directly affects the description of the geometry of the level lines 
of the corresponding potentials.

 Here we will be interested mainly in open (non-closed) level 
lines of potentials $f (x,y)$. The main thing that we would like to 
present here is the division of open level lines into topologically 
regular and chaotic ones. Lines of both the first and second types 
correspond to the general situation and appear in complementary sets 
in the parameter space of the problem. The lines of these two types 
are distinguished by their global geometry, which is relatively simple 
for lines of the first type and rather complex for lines of the second type. 
In addition to the simplicity of the global geometry, topologically regular 
level lines are stable with respect to small variations of the problem 
parameters, while chaotic lines are unstable. As a consequence, the sets 
in the parameter space that correspond to the appearance of level lines 
of different types have a rather nontrivial structure. In the general case, 
the set corresponding to topologically regular level lines consists 
of an infinite number of Stability Zones corresponding to the appearance 
of open level lines with different values of certain topological invariants.
The set corresponding to the appearance of chaotic trajectories 
is the addition to the set of Stability Zones and has a non-trivial 
fractal structure (except for special cases).

\section{Topologically regular and chaotic level lines of 
quasi-periodic potentials}
\setcounter{equation}{0}

\begin{figure}[t]
\begin{center}
\includegraphics[width=\linewidth]{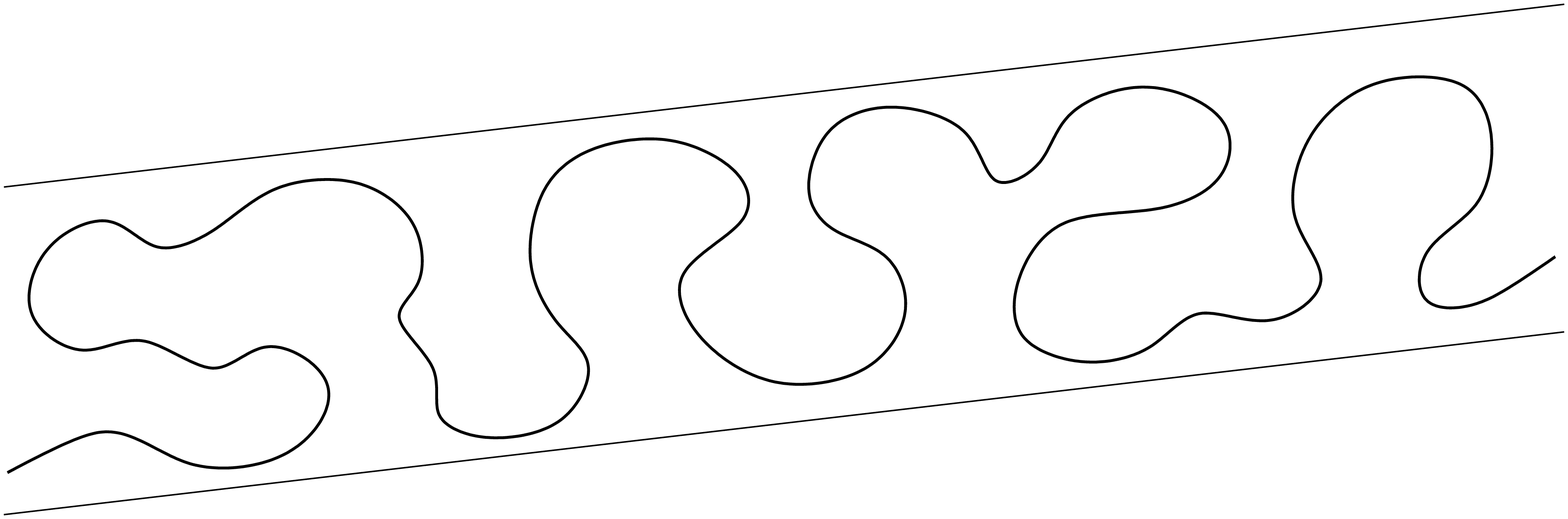}
\end{center}
\caption{Topologically regular level line lying in a straight strip 
of finite width (schematically).}
\label{Fig1}
\end{figure}

\begin{figure}[t]
\begin{center}
\includegraphics[width=\linewidth]{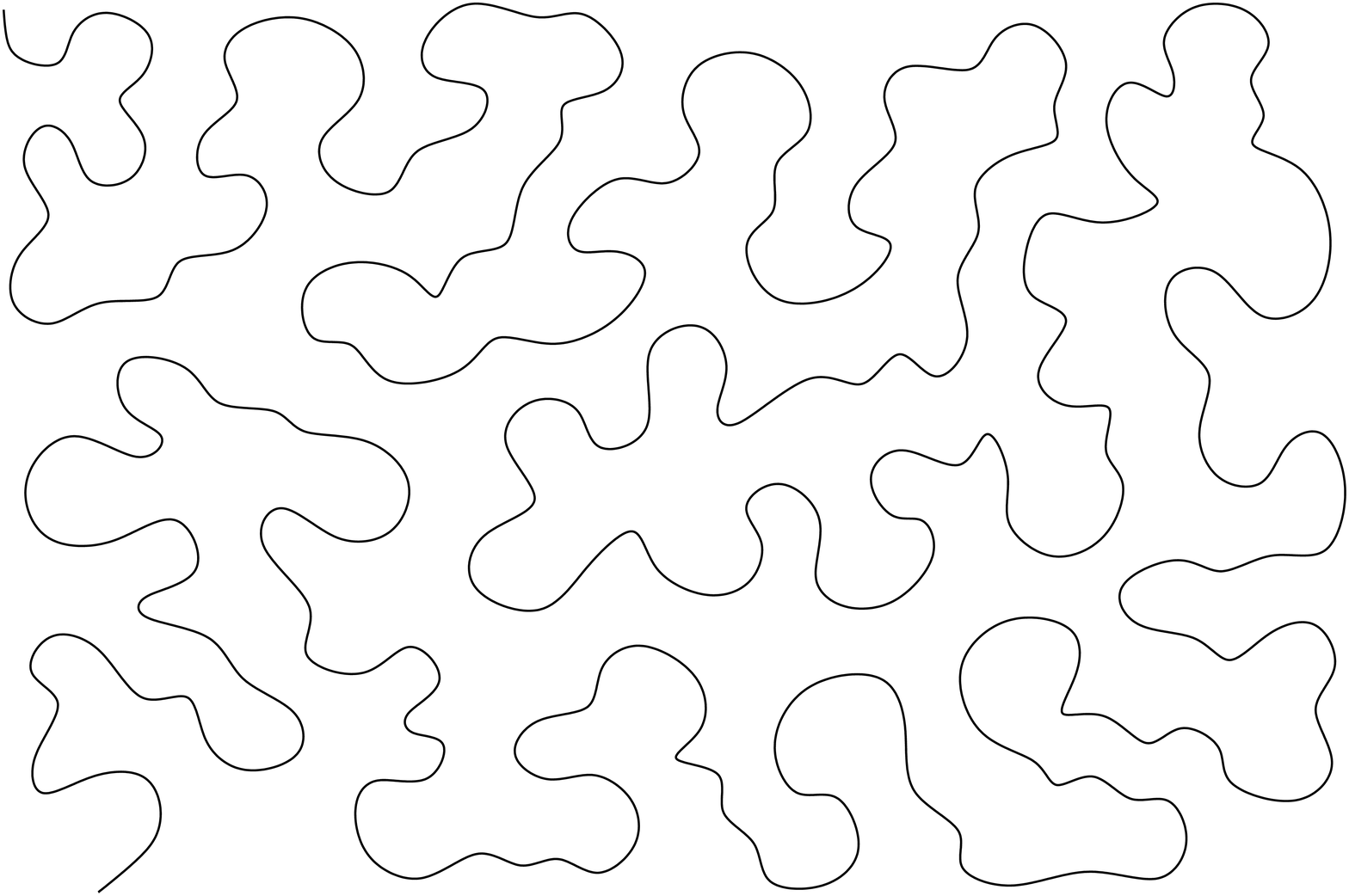}
\end{center}
\caption{A chaotic level line wandering over a plane (schematically).}
\label{Fig2}
\end{figure}

 First of all, we describe here the difference between topologically 
regular and chaotic open level lines of a quasiperiodic function $f (x, y)$.
The main feature of topologically regular open level lines is that each 
such line lies in a straight strip of finite width and passes through it 
(Fig. \ref{Fig1}). Chaotic level lines, on the contrary, cannot be enclosed in 
straight strips of finite width and wander around the plane in some 
pseudo-random way (Fig. \ref{Fig2}). Let us also note here that topologically 
regular level lines are not periodic in the case of general position.

 The second distinguishing feature of topologically regular level lines 
is their stability with respect to small variations in the parameters of 
the problem, while chaotic level lines are unstable with respect to 
arbitrarily small variations in the parameters. 

 The third important feature of topologically regular level lines is 
their connection with topological numbers that define their mean direction.
The topological numbers for the Novikov problem with four quasi-periods 
have the form of (irreducible) integer quadruples 
$(m^{1}, m^{2}, m^{3}, m^{4})$ and, as applied to the situation under 
consideration, are described as follows.

 Let there be a quasi-periodic pattern on the plane, which arises as a 
result of the superposition (possibly with mutual influence) of two 
periodic potentials $V (x, y)$ and $U (x, y)$ with periods 
$({\bf v}_{1}, {\bf v}_{2})$ and $({\bf u}_{1}, {\bf u}_{2})$ 
respectively. Consider in the plane the basis 
$({\bf v}^{\prime}_{1}, {\bf v}^{\prime}_{2})$
reciprocal to the basis $({\bf v}_{1}, {\bf v}_{2})$
and the basis $({\bf u}^{\prime}_{1}, {\bf u}^{\prime}_{2})$
reciprocal to the basis $({\bf u}_{1}, {\bf u}_{2})$
$${\bf v}^{\prime}_{i} \cdot {\bf v}_{j} \, = \, \delta_{ij} \, ,
\quad  {\bf u}^{\prime}_{i} \cdot {\bf u}_{j} \, = \,
\delta_{ij} $$

 Then the mean direction ${\bf l}$ of topologically regular level 
lines of the resulting potential is determined by the relation
\begin{equation}
\label{NumbersPlane}
\left( m^{1} {\bf v}^{\prime}_{1} + m^{2} {\bf v}^{\prime}_{2} + 
m^{3} {\bf u}^{\prime}_{1} + m^{4} {\bf u}^{\prime}_{2} \right) \cdot {\bf l}
\,\,\, = \,\,\, 0 
\end{equation}
with some irreducible integer quadruple $(m^{1}, m^{2}, m^{3}, m^{4})$. 

\begin{figure}[t]
\begin{center}
\includegraphics[width=\linewidth]{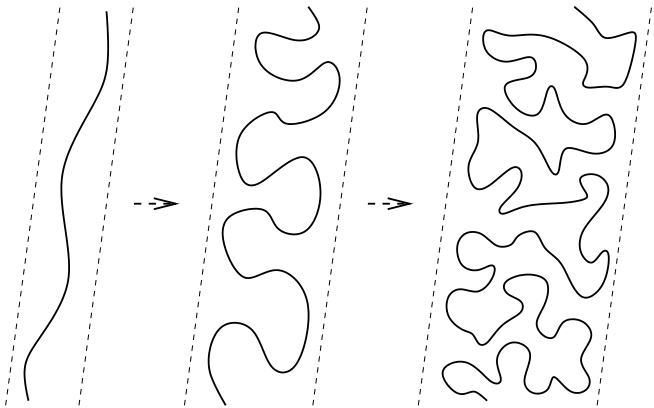}
\end{center}
\caption{Complication of the shape of topologically regular level lines 
with increasing numbers $(m^{1}, m^{2}, m^{3}, m^{4})$ (schematically).}
\label{Fig3}
\end{figure}

 The numbers $(m^{1}, m^{2}, m^{3}, m^{4})$ are locally stable and do not 
change with small variations of parameters of the problem. Thus, the set 
of existence of topologically regular open potential level lines in the 
parameter space is actually divided into Stability Zones, each of which has 
its own values of $(m^{1}, m^{2}, m^{3}, m^{4})$. In general case the 
quadruples  $(m^{1}, m^{2}, m^{3}, m^{4})$ form a certain subset 
in $\mathbb{Z}^{4}$ and the number of Stability Zones in the parameter 
space may be either finite or infinite. We also note here that with 
increasing values of $(m^{1}, m^{2}, m^{3}, m^{4})$ the corresponding level 
lines become more and more complex. In particular, the width of the strips 
containing such level lines increases, and the level lines themselves 
acquire a chaotic shape on small scales (Fig. \ref{Fig3}).

 Among other parameters, topologically regular level lines are also stable 
under small variations of the energy value $f (x, y) = \epsilon$. 
Thus, topologically regular level lines in general position exist in some 
finite connected energy interval. The exception is the level lines that 
appear at the boundaries of the Stability Zones, where this interval 
collapses into a single energy level. In the case of general position 
we have also a finite energy interval, for which the geometry of the 
regions $f (x, y) \leq \epsilon$ is similar to the geometry of 
topologically regular open level lines. The topologically regular open 
level lines are also stable in this case with respect to sufficiently small 
perturbations of the potential $f (x, y)$ of an arbitrary form.

 It is easy to see that the specific properties of topologically 
regular trajectories may manifest themselves, in particular, in transport 
phenomena in the corresponding two-dimensional systems.

 As for the behavior of chaotic level lines, it is actually very diverse.
Let us say here that the behavior of chaotic level lines of quasi-periodic 
potentials has been fairly well studied by now for the Novikov problem 
with three quasi-periods (see, for example, 
\cite{tsarev,DynnBuDA,dynn2,zorich2,zorich3,DeLeoDynnikov1,DeLeoDynnikov2,
Dynnikov2008,Skripchenko1,Skripchenko2,DynnSkrip1,DynnSkrip2,
AvilaHubSkrip1,TrMian}). In this case, the chaotic level lines should 
actually be divided into certain types that have their own specific 
features. In the case of four quasi-periods, the behavior of chaotic 
open level lines is more complex, and is still very far from its complete 
description. Here we will classify as chaotic level lines all non-periodic 
open level lines that do not satisfy the properties listed above 
for topologically regular level lines. It is easy to see that no quasi-periodic 
potential can have simultaneously topologically regular and chaotic open level 
lines. Thus, each of the quasi-periodic potentials can be attributed to one of 
two types in accordance with the geometry of its open level lines.

 In this paper, we describe the picture that arises in the problem 
we are considering only from the most general point of view. A more precise 
description depends on the specific situation and requires additional 
clarification of the conditions of the problem.

\section{The emergence of topologically regular and chaotic level lines 
in the general situation and special cases}
\setcounter{equation}{0}

 It is easy to see that in the Novikov problem with four quasi-periods,
a part of the parameters is given by the parameters of the affine embedding 
of the two-dimensional plane into the four-dimensional space.
These parameters, in turn, include the parameters of the linear map 
$\mathbb{R}^{2} \rightarrow \mathbb{R}^{4}$ as well as the shift of the 
origin for the full affine embedding. Here, in fact, we will be mainly 
interested in the division of open potential level lines into regular and 
chaotic ones, the type of which does not change under affine transformations 
of the plane $\mathbb{R}^{2}$ itself. For us, therefore, the essential parameters 
will be represented by a two-dimensional direction in four-dimensional space, 
i.e. a point of the Grassmann manifold $G_{4,2}$, and the values of shifts 
in one-dimensional directions orthogonal to it. It is easy to see that 
for a fixed two-dimensional direction we have a two-parameter family of 
two-dimensional planes representing essentially different affine embeddings
$\mathbb{R}^{2} \rightarrow \mathbb{R}^{4}$.

 In addition, the Novikov problem naturally contains the space of 4-periodic 
smooth functions, i.e., space of smooth functions on the four-dimensional torus
$\mathbb{T}^{4}$. Let us formulate here the main results obtained so far for the 
Novikov problem with four quasi-periods (\cite{DynNov}).

\vspace{1mm}

 There is an open everywhere dense subset 
$S \subset C^{\infty} (\mathbb{T}^{4})$ 
of 4-periodic functions $F$ and an open everywhere dense subset 
$X_{F} \subset G_{4,2}$, depending on $F$, such that for any $\Pi \in X_{F}$ 
any level $M^{3}_{c} = \{ F = c \} $ contains only stable topologically regular 
level lines of functions $f$ (or does not contain open level lines of $f$).

 In addition, for any regular level line, the width of the strip bounding it, 
as well as the diameter of any compact level line of the functions $f$, are 
bounded from above by one constant $C$, which does not depend on the choice of 
the affine plane (of the direction $\Pi$) containing the level line, provided 
that the level $c$ and direction $\Pi \in X_{F}$ are fixed.

\vspace{1mm}

 The above statements are formulated in our original setting, where the 
function $f (x, y)$ is obtained by restricting of a 4-periodic function 
$F (z^{1}, z^{2}, z^{3}, z^{4})$ under the affine embedding
$\mathbb{R}^{2} \rightarrow \mathbb{R}^{4}$. It is easy to see that the 
level lines of the functions $f(x, y)$ are given then by the intersections 
of the corresponding two-dimensional planes with the periodic three-dimensional 
manifolds $F (z^{1}, z^{2}, z^{3}, z^{4}) = {\rm const}$.

 As we have already said, the mean direction of topologically regular level 
lines of the function $f (x, y)$ is determined by some quadruple of integers
$(m^{1}, m^{2}, m^{3}, m^{4})$. In the general case, assuming that the function 
$F (z^{1}, z^{2}, z^{3}, z^{4})$ is periodic with respect to the standard integer 
lattice, the mean direction of the topologically regular level lines 
of $f (x, y)$ is given by the intersection of the corresponding two-dimensional 
plane with an integral plane in $\mathbb{R}^{4}$ given by the equation
\begin{equation}
\label{NumbersSpace}
m^{1} z^{1} \, + \, m^{2} z^{2} \, + \, m^{3} z^{3} \, + \, m^{4} z^{4} 
\,\, = \,\, 0 
\end{equation}

 The set corresponding to the appearance of topologically regular open level 
lines, therefore, consists of Stability Zones with different values of
$(m^{1}, m^{2}, m^{3}, m^{4})$, the union of which is everywhere dense in the 
described parameter space. In the general case, the number of such Zones is 
infinite, and the corresponding numbers $(m^{1}, m^{2}, m^{3}, m^{4})$ can be 
arbitrarily large. 

 Here we would like to pay special attention to the fact that in the case 
of the presence of topologically regular open level lines, 
the above statements apply immediately to the entire family of parallel 
two-dimensional planes embedded in four-dimensional space. Thus, in the case 
of the appearance of such level lines, we immediately obtain 
their description for a whole family of periodic potentials corresponding to 
all embeddings of the same direction.

 It should also be added here that, as follows from the results 
of \cite{NovKvazFunc,DynNov}, whenever the potentials we consider are not 
periodic, the mean directions of their topologically regular level lines are the 
same in all indicated planes. At the same time, if such potentials are periodic, 
the mean direction of their open level lines can be different in different 
planes of the family (but not necessarily).

 In addition, the open level lines of every function $f (x, y)$
appear in some connected energy interval $[\epsilon_{1}, \epsilon_{2}]$
(which can shrink to a single point $\epsilon_{0}$). When topologically regular 
open level lines appear at a non-periodic potential, the corresponding interval 
also coincides for all planes of the family we have indicated. At the same time, 
if such potentials are periodic, these intervals may differ in different planes 
of this family.

 It is easy to see that all open level lines of periodic potentials 
are periodic, and these potentials themselves are not generic potentials. 
In our situation here, such potentials will appear rather as exceptional 
potentials.

 It also follows from the above statements that chaotic level lines cannot be 
stable in the described parameter space, since each such situation is a point 
of accumulation of Stability Zones corresponding to the presence of topologically 
regular open level lines. As we noted above, the corresponding topologically 
regular level lines in this situation acquire chaotic properties on finite scales.

\vspace{1mm}

 The above statements describe the structure of the set of appearance 
of regular level lines in the most general parameter space corresponding 
to the Novikov problem with four quasi-periods. When considering a specific 
problem, we must additionally consider the embedding of the parameter space 
of this problem into the complete parameter space described above.

 In our case, we are dealing with the superimposition of two periodic 
pictures on top of each other. It is natural to assume that the position 
of one of the pictures is fixed. The parameters of the problem are then 
two independent shifts and a rotation of the second picture relative 
to the first. 

 Denoting the corresponding Euclidean transformation of the 
plane by $A (x,y)$
$$A (x, y) \,\,\, = \,\,\, \left(
\begin{array}{cc}
\cos \alpha  &  \sin \alpha  \\
- \sin \alpha  &  \cos \alpha 
\end{array}  \right)  
\left( 
\begin{array}{c}
x  \\
y
\end{array} \right)
\,\, + \,\, \left(
\begin{array}{c}
a_{1}  \\
a_{2} 
\end{array}  \right)
\,\,\, , $$
we can define the embedding 
$\mathbb{R}^{2} \rightarrow \mathbb{R}^{4}$
according to the formula
\begin{equation}
\label{Vlozhenie}
(x, y) \quad \rightarrow \quad \left( x, y, A (x,y) \right) 
\end{equation}

 Thus, the transformation $A (x, y)$ will play a double role for us.
On the one hand, it determines the shift of one of the periodic potentials 
relative to the other in the plane. On the other hand, 
we will also use it to define an embedding 
$\mathbb{R}^{2} \rightarrow \mathbb{R}^{4}$
in order to define the resulting potential as the restriction of some 
4-periodic function $F (z^{1}, z^{2}, z^{3}, z^{4})$ to the embedded plane. 
As we have already said, we assume here that one periodic potential 
$V (x, y)$ is stationary in the plane, while the second potential $U (x, y)$ 
undergoes some rotation and shift before being superimposed.

 For all the transformations $A (x, y)$ the projection of the embedding 
(\ref{Vlozhenie}) on the plane $(z^{1}, z^{2})$ is the identical transformation, 
while the parameters of the transformation $A (x,y)$ determine the shifts and 
rotation of the projection on the plane $(z^{3}, z^{4})$. As it is easy to see,
the rotation in the transformation $A (x, y)$ corresponds to the rotation  
of this projection while the shifts in $A (x, y)$ correspond to parallel 
shifts of the nested plane in the four-dimensional space. It is also easy to 
see that for any fixed value of the rotation angle in the transformation 
$A (x, y)$ all such shifts fill the entire four-dimensional space.

 It is easy to see that if the resulting potential $f (x,y)$ in the plane 
is given by the simple sum of periodic potentials 
$f_{1} (x,y) = V (x, y)$ and $f_{2} (x,y) = U (A (x,y))$, 
the corresponding periodic function 
$F (z^{1}, z^{2}, z^{3}, z^{4})$ in four-dimensional space can be defined as
$$F (z^{1}, z^{2}, z^{3}, z^{4}) \,\,\, = \,\,\, 
V (z^{1}, z^{2}) \,\, + \,\, U (z^{3}, z^{4}) $$

 As we can see, the function $F (z^{1}, z^{2}, z^{3}, z^{4})$
is fixed in this case and the problem parameters affect only the parameters
of the embedding $\mathbb{R}^{2} \rightarrow \mathbb{R}^{4}$. The same 
situation actually takes place also if the resulting potential $f (x,y)$
is a local function of the values $f_{1} (x,y)$ and $f_{2} (x,y)$
taken at the same point $(x,y)$: $f = Q (f_{1}, f_{2})$. In this case,  
the function $F (z^{1}, z^{2}, z^{3}, z^{4})$ is given by the 
expression
$$F (z^{1}, z^{2}, z^{3}, z^{4}) \,\,\, = \,\,\, 
Q \left( V (z^{1}, z^{2}), U (z^{3}, z^{4}) \right) $$

 In both cases, the function $F (z^{1}, z^{2}, z^{3}, z^{4})$ is 
obviously periodic in $\mathbb{R}^{4}$, with periods
\begin{equation}
\label{Periods}
({\bf v}_{1}, 0, 0) , \,\,\, ({\bf v}_{2}, 0, 0) , \,\,\,
(0, 0, {\bf u}_{1}) , \,\,\, (0, 0, {\bf u}_{2}) , 
\end{equation}
where $({\bf v}_{1}, {\bf v}_{2})$, $({\bf u}_{1}, {\bf u}_{2})$ 
represent the periods of the potentials $V (x, y)$ and $U (x, y)$ 
respectively.

 The most general case is when the resulting potential $f (x, y)$
is a non-local functional of potentials $f_{1} (x, y)$ and $f_{2} (x, y)$
\begin{equation}
\label{Functional}
f (x, y) \,\, = \,\, Q [f_{1}, f_{2}] (x,y) 
\end{equation}
(for example, due to local deformation of atomic lattices as a result of their 
interaction). For translationally invariant functionals
$f [f_{1}, f_{2}]$ the corresponding function $f (x, y)$ will also be a 
quasi-periodic function with four quasi-periods. The corresponding function 
$F (z^{1}, z^{2}, z^{3}, z^{4})$ for a given transformation $A$ is defined
as follows.

 Let us fix the angle $\alpha$ in the transformation $A$ and consider
all the transformations $A^{\prime}$ with the same $\alpha$ and all
possible values of $a_{1}$ and $a_{2}$. Each transformation $A^{\prime}$
defines a certain configuration of superposition of the potentials
$V (x, y)$ and $U (x, y)$ in the plane, and thus determines the 
corresponding values of the functional $f (x, y, a_{1}, a_{2})$ 
in it. On the other hand, each transformation $A^{\prime}$ defines 
an embedding of the plane in $\mathbb{R}^{4}$ and, as we have already 
said, the complete family of corresponding parallel planes fills the 
entire four-dimensional space. The latter circumstance allows us to 
uniquely define the coordinates $z^{i}$ as (linear) functions 
of $x$, $y$, $a_{1}$, and $a_{2}$
$$z^{i} \,\,\, = \,\,\, z^{i} (x, y, a_{1}, a_{2}) $$ 
and vice versa
$$x = x ({\bf z}) , \,\,\, y = y ({\bf z}) , \,\,\,
a_{1} = a_{1} ({\bf z}) , \,\,\, a_{2} = a_{2} ({\bf z}) $$

 The function $F (z^{1}, z^{2}, z^{3}, z^{4})$ is then naturally 
defined as
\begin{equation}
\label{FunctionF}
F ({\bf z}) \,\, = \,\, f \left( 
x ({\bf z}), y ({\bf z}), a_{1} ({\bf z}), a_{2} ({\bf z}) \right) 
\end{equation}

 It is not difficult to verify that the function $F ({\bf z})$
constructed in this way has the same periods (\ref{Periods})
as in the previous cases. Indeed, returning to the formula 
(\ref{Vlozhenie}), we see that the shifts
$$(z^{1}, z^{2}) \, \rightarrow \, (z^{1}, z^{2}) + {\bf v}_{i} \, ,
\,\,\, (z^{3}, z^{4}) \, \rightarrow \, (z^{3}, z^{4}) $$
correspond to the transformation of the parameters
$(x, y, a_{1}, a_{2})$ such that
$$(x, y) \, \rightarrow \, (x, y) + {\bf v}_{i} \, ,  \,\,\,
A^{\prime} (x, y) \, \rightarrow \, A^{\prime} (x, y) $$

 In the same way, the shifts
$$(z^{1}, z^{2}) \, \rightarrow \, (z^{1}, z^{2}) \, ,
\,\,\, (z^{3}, z^{4}) \, \rightarrow \, (z^{3}, z^{4}) + {\bf u}_{i} $$
correspond to
$$(x, y) \, \rightarrow \, (x, y) \, ,  \,\,\,
A^{\prime} (x, y) \, \rightarrow \, A^{\prime} (x, y) + {\bf u}_{i} $$

 It is easy to see, therefore, that in both cases the initial potentials 
$f_{1} (x,y) = V (x, y)$ and $f_{2} (x,y) = U (A^{\prime} (x,y))$
in the plane remain unchanged, which also entails the invariance of the 
functional (\ref{FunctionF}). 

 Unlike the previous cases, however, the form of the function $F ({\bf z})$
depends here on the value of the angle $\alpha$ in the transformation
$A (x, y)$. Thus, we see that in the most general case, the parameters 
of our problem are mapped to the most complete space of parameters of 
the Novikov problem, including the space of periodic functions 
$F ({\bf z})$ in $\mathbb{R}^{4}$ (with fixed periods).

 Let us say here that we use the term potentials rather conditionally, 
and the functions $V (x,y)$, $U (x, y)$ and $f (x,y)$ may  actually be any 
functionals that describe the properties of the system under consideration.
Also, the dependence (\ref{Functional}) may be a very complex functional 
dependence, depending on the type of functionals $V (x,y)$, $U (x, y)$ 
and $f (x,y)$. In any case, however, all the assertions formulated above 
will hold. Let us note also, that if we introduce the new coordinate
system $(z^{\prime 1}, z^{\prime 2}, z^{\prime 3}, z^{\prime 4})$,
where the periods of the function $F ({\bf z})$ play the role of the 
basis vectors, we will directly obtain the relation (\ref{NumbersPlane})
from the relation (\ref{NumbersSpace}). 

\vspace{1mm}

\begin{figure}[t]
\begin{center}
\includegraphics[width=\linewidth]{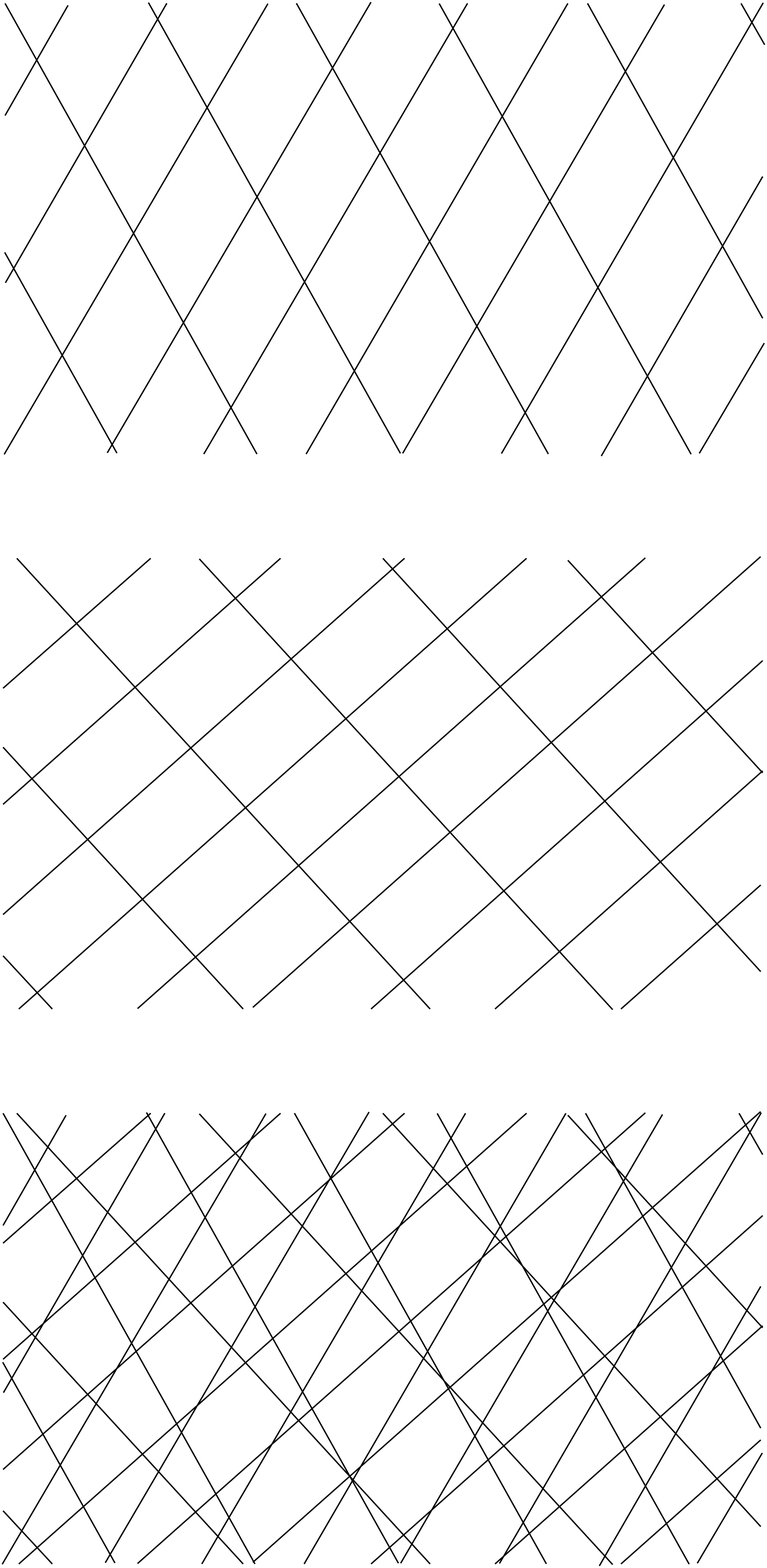}
\end{center}
\caption{Superposition of two periodic potentials of the most 
general form (schematically).}
\label{Fig4}
\end{figure}

 To describe the generic situation in the problem considered here, 
we can, for example, impose the condition that the potentials 
$V (x, y)$ and $U (x, y)$ do not have nontrivial rotational symmetry, 
and the lengths of the periods ${\bf v}_{1}$, ${\bf v}_{2}$,
${\bf u}_{1}$, ${\bf u}_{2}$ are linearly independent over the field 
of rational numbers (Fig. \ref{Fig4}). In this formulation, in particular, 
periodic potentials do not arise at all in any superposition of the layers 
and the only parameter that specifies the type of open level lines is 
the rotation angle $\alpha$.

 For generic mappings of the parameters of our problem to the 
above-described space of parameters of the general Novikov problem,
we must observe in general the intersection of the image 
of a mapping both with the Stability Zones and with their 
complement in this space. In the generic situation, we must 
therefore observe the separation of the cases of topologically 
regular and chaotic open level lines of emerging potentials 
depending on the angle $\alpha$. The set of angles $\alpha$
corresponding to the topologically regular case represents 
then a (finite or infinite) set of intervals on the unit circle, 
each of which corresponds to its own values of
$(m^{1}, m^{2}, m^{3}, m^{4})$. The set of angles corresponding 
to the chaotic case is the addition to the union of such intervals 
and may have a fractal structure in the most general situation.
On the whole, such a situation should arise for many systems 
satisfying the above conditions of the general position.

 In fact, a similar situation should apparently also arise 
under weaker assumptions about the geometry of the potentials
$V (x, y)$ and $U (x, y)$. Most likely, to observe such a situation, 
it is sufficient to require only the absence of rotational symmetries 
of the same order for the superimposed potentials. Pairs of 
corresponding materials also appear in many interesting 
systems. We would like to give here just one example of such a pair, 
which has been of great interest in recent times (see \cite{Akamatsu}).
It should also be added here that in special cases when the periods 
of $V (x, y)$ and $U (x, y)$ are commensurate, in addition to the sets 
described above, special angles $\alpha$ may also arise for which the 
resulting potential is purely periodic.

 It is easy to see, however, that many important and interesting 
physical two-layer systems are not systems in general position in 
the sense described above. Many of them, in particular, have a nontrivial 
rotational symmetry common to both potentials $V (x, y)$ and $U (x, y)$. 
Probably the most important example of such a system is two-layer graphene, 
with layers arranged at arbitrary angles to each other. Both layers in this 
case are identical to each other and have sixth-order rotational symmetry.
For almost all rotation angles $\alpha$ the resulting potential 
is quasi-periodic, and only for some (magic) angles special periodic 
potentials appear (see for example \cite{Shallcross1,Shallcross2}).

\begin{figure}[t]
\begin{center}
\includegraphics[width=\linewidth]{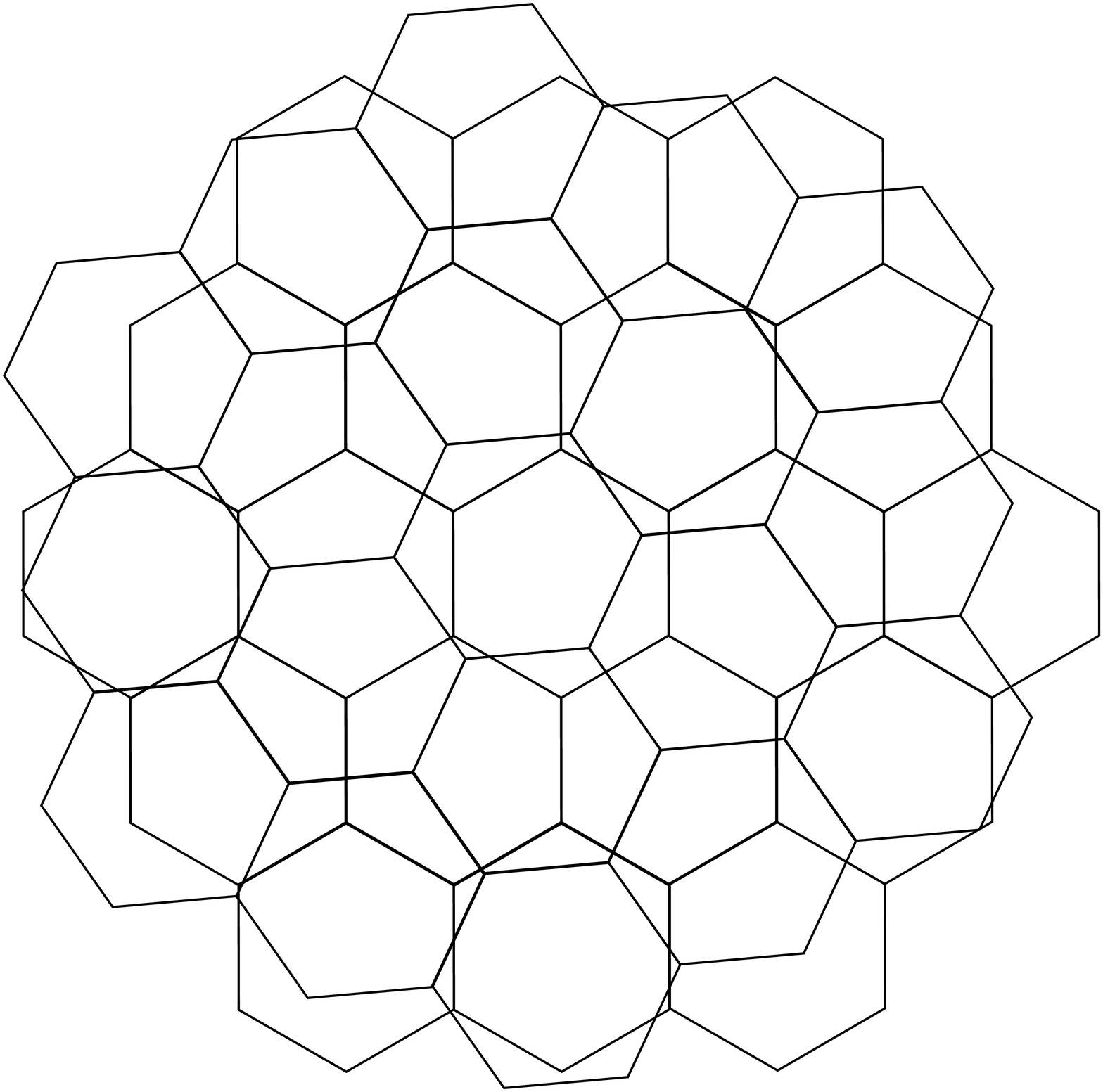}
\end{center}
\caption{Two identical potentials of hexagonal symmetry, 
superimposed in a plane with rotation through an arbitrary 
angle (schematically).}
\label{Fig5}
\end{figure}

 It is easy to see that in the case of quasi-periodic (non-periodic)
potentials we cannot observe here topologically regular open level lines.
Indeed, topologically regular level lines must exist for potentials 
in the entire family of planes of a given direction in $\mathbb{R}^{4}$
(i.e., for all values of parameters $a_{1}$ and $a_{2}$). On the other hand, 
by choosing the parameters $a_{1}$ and $a_{2}$ in the right way, 
we can achieve the occurrence of a quasi-periodic potential
which has rotational symmetry. It is easy to see then that for such 
a potential the appearance of topologically regular level lines is 
impossible.

 It can be seen, therefore, that for almost all angles $\alpha$
the open level lines in two-layer graphene belong to the class 
of chaotic ones. The only exceptions are magic angles corresponding 
to the appearance of periodic potentials in the plane. In this case, 
the overall picture in the plane also depends on the shift parameters
$a_{1}$ and $a_{2}$. Depending on their values, we can observe here 
either (unstable) open periodic level lines of different directions 
for different $a_{1}$ and $a_{2}$, or periodic nets of singular trajectories 
that appear only at a single energy level (for example, for periodic 
potentials with rotational symmetry).

 Another important example of a nongeneric situation is when the 
superimposed potentials have rotational symmetry of the same order, 
however, they are not identical to each other. An example of this may 
be given by the system formed by graphene aligned with hexagonal 
boron nitride, where both layers have hexagonal symmetry, but the 
periods of the corresponding potentials are not exactly equal to each other
(let us quote here just a few of the many works devoted to this 
important system 
\cite{Dean,Abanin,YankowitzXueCormode,Geim,YankowitzXueLeRoy,TitovKatsnelson}).
The chaotic level lines arising in such a system can indeed have a very 
complex shape. We would like to especially refer here to the work 
\cite{TitovKatsnelson} where the shape of such trajectories was 
studied in the percolation theory approximation. We also note here that 
the paper \cite{TitovKatsnelson} considers the situation when the 
superimposed potentials have a very strong interaction with each other, 
and the investigated functional is a complex functional of the resulting 
potential, being a parameter of the electronic spectrum in this system
(averaged electron mass). The problem studied in the work 
\cite{TitovKatsnelson} is directly related to the description of the 
transport electronic properties in the corresponding two-dimensional 
system.

\section{Conclusions}
\setcounter{equation}{0}

 We consider the geometry of open level lines of special 
potentials obtained as a result of a superposition of periodic 
potentials on the plane. The consideration is based on the connection 
of such potentials with the theory of quasi-periodic functions and, 
in particular, with the Novikov problem for potentials with four 
quasi-periods on the plane. As follows from the general results, 
each of these potentials has either topologically regular 
or chaotic open level lines that replace each other in 
a rather nontrivial way as the parameters of the potential change.
At the same time, the paper considers both families of potentials 
that correspond to the most general situation, and special families 
that are important from the physical point of view.

\end{document}